\documentclass[12pt]{iopart}
\usepackage{graphicx}
\usepackage{epsfig}
\expandafter\let\csname equation*\endcsname\relax
\expandafter\let\csname endequation*\endcsname\relax
\usepackage{amsmath}
\usepackage{iopams}

\newcommand{\bra}[1]{\ensuremath{\langle{#1}|\,}}
\newcommand{\ket}[1]{\ensuremath{\,|{#1}\rangle}}
\newcommand{\braket}[1]{\ensuremath{\langle{#1}\rangle}}

\begin{document}

\title
{Superoperator coupled cluster method for nonequilibrium density matrix}

\author{Alan A. Dzhioev$^1$ and D. S. Kosov$^2$}

\address{$^1$ Bogoliubov Laboratory of Theoretical Physics, Joint Institute for Nuclear Research,  RU-141980 Dubna, Russia}

\address{$^2$ School of Engineering and Physical Sciences, James Cook University, Townsville, QLD, 4811, Australia }


\begin{abstract}
We develop superoperator  coupled cluster  method for nonequilibrium open
many-body quantum systems described by  Lindblad master equation. The method is universal and applicable to systems of interacting fermions, bosons or their mixtures.
We present a general theory and consider its application to the problem of quantum transport through the system with electron-phonon correlations. The results are assessed against the perturbation theory and nonequilibrium configuration interaction theory calculations.
\end{abstract}

\date{\today}

\maketitle

\section{Introduction}


The coupled cluster method has proved to be extremely useful in a wide variety of many body calculations ranging from nuclear physics to quantum chemistry \cite{ccm04,Kummel78,bartlett:291}.
In this method the wave function of a quantum many-body system is decomposed in terms of amplitudes for excited "clusters" of finite number of quasiparticles or particle-hole pairs. Coupled cluster theory works for systems of both bosons and fermions and gives high-precision results for the equilibrium properties.  Today, it  is one of the most widely used methods for equilibrium many-body calculations \cite{bartlett:291}.
In this paper we propose the extension of the  coupled cluster method for nonequilibrium open quantum systems governed by the Lindblad master equation.


The paper is organised as follows. Section 2 describes   the Liouville-Fock space and superoperator representation of the Lindblad master equation. Section 3 discusses the
nonequilibrium coupled cluster (NECC)  expansion for the steady-state density matrix of a correlated many-body system. In section 4 we apply the method to the problem of charge transport through the system with electron-phonon
interaction and compare the numerical results with other methods.   Section 5 summarises the main results of the paper. Appendix~A contains some technical details concerning the choice of a reference state for transport calculations. In Appendix~B we give explicit equations for the cluster amplitudes.

\section{Lindblad equation in Liouville-Fock space}

We consider
a correlated quantum system described by the Hamiltonian $H$.
The system is open and exchange energy and particles with the environment.
 We assume that
the reduced density matrix of the system satisfies the Lindblad master equation\cite{Petruccione}:
\begin{align}\label{lindblad}
& i \frac{\partial\rho(t)}{\partial t} =[H,\rho(t)] + i{\Pi}\rho(t),
\end{align}
where ${\Pi}\rho(t)$ is the non-Hermitian dissipator  given by the standard Lindblad form
\begin{align}\label{non_herm}
{\Pi}\rho(t) = \sum_{k} \bigl( 2 L_k\rho(t) L^\dag_k - \{L^\dag_k  L_k,\rho(t) \}\bigr),
\end{align}
and  $L_k$ are so-called Lindblad  operators, which model the effects of the environment~\cite{Petruccione}.
In what follows we consider that $H$ and $L_{k}$ act in the Fock space of the system under consideration, so they can be written in terms of creation and annihilation operators.
The Lindblad master equation~\eqref{lindblad}
describes the time evolution of an open system preserving Hermiticity, normalization, and positivity of the reduced density matrix.

The Lindblad master equation can be converted to a non-Hermitian Schr\"{o}dinger-like form with the use of superoperator formalism~\cite{schmutz78,prosen08,dzhioev11a,dzhioev12}.
Within the formalism, every  Fock space operator $A$ is considered as a super-vector $\ket{A}$ in the Liouville-Fock space.
In particular, if $|m)$ and $|n)$ are vectors in the Fock space, then $|mn)\equiv\ket{|m)(n|}$ is a super-vector in the Liouville-Fock space.
The scalar product in
the Liouville-Fock space is defined as $\braket{A_1|A_2}=\mathrm{Tr}[A^\dag_1 A_2]$.

 In the  Liouville-Fock space we can introduce creation and annihilations superoperators
$\hat a^\dag,\,\widetilde a^\dag$ and $\hat a,\,\widetilde a$:
 \begin{align}\label{super_a}
     &\hat a_k\ket{mn} = \ket{a_k |m)(n|},~~  \widetilde a_k\ket{mn} =\tau_{mn}\ket{|m)(n| a^\dag_k},
  \notag\\
 &\hat a^\dag_k\ket{mn} = \ket{a^\dag_k|m)(n|}, ~~  \widetilde a^\dag_k\ket{mn} =\tau_{mn}\ket{|m)(n| a_k},
\end{align}
  where $|m)=|m_1,m_2,\ldots)$ are eigenvectors of the particle number operator, i.e., $a^\dag_k a_k|m)=m_k |m)$. The phase $\tau_{mn}=1$ for bosonic creation and annihilation superoperator,
 while  for fermionic ones we have $\tau_{mn}=i(-1)^\mu$, where $\mu=\sum_k(m_k+n_k)$.~\footnote{Such defined creation and annihilation superoperators are Hermitian conjugate to each other, $\hat a^\dag= (\hat a)^\dag$,
  $\widetilde a^\dag= (\widetilde a)^\dag$, and satisfy the same (anti)commutation relations as their Fock space counterparts, $[\hat a,\hat a^\dag]_\mp=1$ etc. See~\cite{dzhioev12} for more details.}
For an operator in the Fock space $A=A(a^\dag,a)$ we formally define two superoperators $\hat A = A(\hat a^\dag, \hat a)$, $\widetilde A = A^*(\widetilde a^\dag, \widetilde a)$ and refer to them
as non-tilde and tilde superoperators, respectively. Then, we can find that the respective super-vector $\ket{A}$ is given by
 \begin{equation}\label{ketA}
  \ket{A} = \hat A\ket{I} = \sigma_A \widetilde A^\dag\ket{I},
\end{equation}
where the super-vector $\ket{I}$ corresponds to the Fock space identity operator, and the phase  $\sigma_A = -i(+1)$ if $A$ is a fermionic (bosonic) operator.
Moreover, for the product of operators the following relations hold:
 \begin{equation}\label{A1A2}
   \ket{A_1A_2}=\hat A_1\ket{A_2}=\tau\widetilde A^\dag_2\ket{A_1}.
 \end{equation}
Here $\tau=i$ if both $A_1$ and $A_2$ are fermionic and $\tau=\sigma_{A_2}$ otherwise. The  Liouville-Fock space is  the physical name for  what is known in mathematics as CAR/CCR algebra of observables which admit the Gelfand-Naimark-Segal (GNS) representation, and the Liouville superoperator defined below is the generator of dynamics in this larger space.

Using the correspondence between Fock space operators and Liouville-Fock space super-vectors and relations~\eqref{A1A2}, we rewrite
the Lindblad master equation~\eqref{lindblad} as a Schr\"{o}dinger-like equation for the density matrix
 \begin{equation}\label{Schrodinger}
   i\frac{\partial}{\partial t}\ket{\rho(t)}  =L\ket{\rho(t)}.
\end{equation}
Here the superoperator $L$ (Liouvillian) is given by
\begin{equation}\label{LSchrodinger}
  L=\hat H - \widetilde H - i\sum_{k}\Pi_k,
\end{equation}
and the non-Hermitian dissipators $\Pi_k$ read as
\begin{equation}
\Pi_{k} =\hat L_{k}^\dag \hat L_{k} + \widetilde L_{k}^\dag \widetilde L_{k} - 2\sigma_{L_k}\hat L_{k} \widetilde L_{k}.
\end{equation}
The statistical average of any operator is
\begin{equation}\label{average}
  \braket{A}=\mathrm{Tr}[A\rho(t)]=\braket{I|A\rho(t)}=\braket{I|\hat A|\rho(t)}.
\end{equation}

Before proceeding further let us mention some important properties of the Liouvillian and introduce terminology that will be used throughout the paper.
The connection between non-tilde and tilde superoperators is given by the tilde-conjugation rules
 \begin{align}\label{TCR}
 (c_1\hat A_1 + c_2\hat A_2)\widetilde{}=c_1^*\widetilde A_1 + c_2^*\widetilde A_2,~~~(\hat A_1\hat A_2) \widetilde{} = \widetilde A_1\widetilde A_2,~~~(\widetilde A)\widetilde{}=\hat A.
 \end{align}
By applying tilde-conjugation
 to Eq.~\eqref{Schrodinger} we get $(L)\widetilde{}=-L$.
The super-vector is called tilde-invariant if  $\ket{A}\widetilde{}\equiv\widetilde A\ket{I}=\ket{A}$. The examples of tilde-invariant super-vectors
are $\ket{I}$ and $\ket{\rho(t)}$. Taking the time derivative of the normalization condition
\begin{equation}\label{norm}
  \braket{I|\rho(t)}=\mathop\mathrm{Tr}[\rho(t)]=1,
\end{equation}
we find $\bra{I}L=0$, i.e., $\bra{I}$ is the left zero-eigenvalue eigenvector of $L$.

\section{Coupled cluster expansion of density matrix}

Consider the problem of obtaining the steady-state  solution of equation~\eqref{Schrodinger}.
We wish to find the density matrix, which corresponds to the right zero-eigenvalue eigenstate of the Liouvillian
\begin{equation}
\label{StSt_def}
L \ket{\rho} =0.
\end{equation}
The problem is much more complicated than in equilibrium case, since the Liouvillian $L$ is non-Hermitian and its spectrum is not bound from below, so that the  zero-eigenvalue eigenvector can not  be found variationally.

If the system is quasi-free, i.e., there are no two-body interactions, the Liouvillian is a non-Hermitian quadratic form which can be readily diagonalized by non-unitary, canonical  transformations:
\begin{equation}
L_0 = \sum_n (\Omega_n \hat\xi^\dag_n\hat \xi_n - \Omega^*_n \widetilde \xi^\dag_n \widetilde \xi_n ).
\end{equation}
Here $\hat\xi$, $\widetilde\xi$ describe the normal modes of quadratic Liouvillian (so-called nonequilibrium quasiparticles). The creation and annihilation superoperators
for nonequilibrium quasiparticle are not mutually Hermitian conjugate but, nevertheless, they satisfy canonical (anti)commutation relations \cite{dzhioev11a,dzhioev12}.
With the use of nonequilibrium quasiparticles, the uncorrelated steady-state density matrix
\begin{equation}
L_0 \ket{\rho_0} =0
\end{equation}
can be  defined as a vacuum of nonequilibrium quasiparticles: $\hat\xi_n  \ket{\rho_0} = \widetilde \xi_n  \ket{\rho_0} =0$.
Note that the bra super-vector $\bra{I}$ plays the role of the right vacuum, i.e.,  $\bra{I}\hat\xi^\dag_n = \bra{I}\widetilde\xi^\dag_n$ =0.
Besides, the right and left vacuums are normalized as $\braket{I|\rho_0}=1$.

We now consider how correlations induced by interaction between nonequilibrium quasiparticles modify this steady-state vacuum. Likewise to the equilibrium case, where correlations produce particle-hole or quasiparticle excitations over uncorrelated ground state, the correlations produce nonequilibrium multiquasiparticle excitations over the reference state $\ket{\rho_0}$.
By analogy with the coupled cluster approach widely used in equilibrium many-body calculations,  we propose
an exponential ansatz for the correlated nonequilibrium steady-state density matrix
\begin{equation}
  \ket{\rho} = e^S\ket{\rho_0},
\end{equation}
where the so-called cluster correlation superoperator $S$ is constructed as a linear combination
of multiconfigurational creation superoperators:
\begin{equation}\label{S}
  S=\sum_i s_i C^\dag_i,
\end{equation}
and  $C^\dag_i$ is a product of creation superoperators $\hat\xi^\dag$, $\widetilde\xi^\dag$.
The expansion coefficients $s_i$ are called cluster amplitudes.
Because the density matrix is tilde invariant, the cluster correlation superoperator  $S$ should be invariant with respect to tilde-conjugation, i.e., $(S)\widetilde{}=S$.
Also note that from $\bra{I}C^\dag_i=0$ it follows that  the normalization condition~\eqref{norm}
is automatically fulfilled for the  ansatz.

With the exponential ansatz for the density matrix, the exact equation~\eqref{StSt_def}
is then rewritten in the similarity-transformed form
\begin{equation}\label{CC_eq}
  {e}^{-S}L{e}^S\ket{\rho_0}=0.
\end{equation}
By taking a scalar product of this equation with the complete set of states $\{C^\dag_j\ket{I},~\forall j\}$ and applying
the nested commutator expansion, we obtain a set of coupled multinomial equations for the cluster amplitudes $s_i$:
\begin{equation}
\label{series}
\bra{I} C_j   \Bigl\{ L + [L,S] + \frac{1}{2!}[[L,S],S] + \ldots \Bigr\} \ket{\rho_0}=0,~~~\forall j.
\end{equation}
Performing the commutations leads to contractions between creation and annihilation superoperators.
Considering that $C^\dag_i$ are formed by creation superoperators, the only
possible contractions occur between annihilation superoperators in the Liouvillian~$L$ and creation superoperators in~$S$.
Following equilibrium coupled cluster theory we call each such contraction as a "link"~\cite{Kummel78}.
Thus, we can say that each element of $S$ in parametrization~\eqref{S} is linked or coupled directly to~$L$.
Moreover, since $L$ contains  a finite number of annihilation superoperators and each commutator removes
one annihilation superoperators, infinite series of commutators in  Eq.~\eqref{series} terminates at a finite order. As a result we obtain
finite-order polynomial equations for the cluster amplitudes $s_i$. Furthermore, since $S$ is tilde-invariant we have $s_i=s^*_j$
for $C^\dag_i=(C^\dag_j)\widetilde{}$.

The  NECC method would be  exact  if all possible  multiconfigurational creation superoperators were included in expansion~\eqref{S}.
 In any real many-body calculations this is usually impossible to achieve
 and we should truncate the expansion~\eqref{S} at some lower order. As a result we get an approximate
nonequilibrium steady-state density matrix.

In the end of this section we  note that there is a certain degree of flexibility in defining the uncorrelated reference state  $\ket{\rho_0}$ for subsequent coupled cluster expansion.
Depending on the problem  it is sometimes convenient  to define normal modes   $\hat \xi $  and $\widetilde \xi $ by diagonalizing only the part of $L_0$ and consider the rest of the $L_0$ as
a perturbation which induces "correlations" in the coupled cluster density matrix expansion.

\section{Application of the nonequilibrium coupled cluster method: charge transport through a vibrational electronic level}

\subsection{Derivations of main equations}

As an application of the NECC method we consider the problem of charge transport through a correlated quantum system (central region)
attached to two macroscopic electrodes, left ($L$) and right ($R$).
The electrodes are modeled as infinite noninteracting electron reservoirs which  have the same temperature $T$ but different chemical potentials $\mu_{L,R}$ through an applied
voltage $V=\mu_L-\mu_R$.
The coupling between the central region  and electrodes is described by the tunnelling Hamiltonian. For the central region, we consider
one electronic level of energy $\varepsilon_0$ coupled linearly to a vibrational mode (phonon) of frequency $\omega_0$:
\begin{equation}\label{H_lHm}
  H_S = \varepsilon_0 \alpha^\dag \alpha + \omega_0 d^\dag d + \kappa \alpha^\dag \alpha(d^\dag + d),
\end{equation}
where $\alpha^\dag$ ($\alpha$) and  $d^\dag~(d)$ are electron and phonon creation (annihilation) operators, respectively. The problem is to
compute the steady-state current trough the central region.

\begin{figure}[t!]
\begin{center}
\includegraphics[width=0.7\columnwidth]{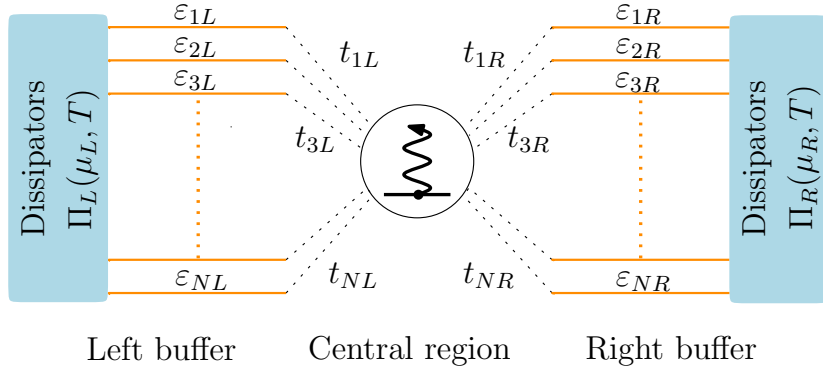}
\end{center}
\caption{Schematic illustration of embedding of an open quantum system for the charge transport problem.  The left and right electrodes are  divided into  an infinite environment  and a finite buffer.
After projecting out the environment degrees of freedom, their effect on the embedded system is represented by the Lindblad dissipators connected to  discrete  energy levels
$\varepsilon_{k\alpha}~(\alpha=L,R)$ of the buffers. The coupling between the central region and the buffers is given by the tunneling matrix elements $t_{k\alpha}$.}
\label{system}
\end{figure}

In~\cite{dzhioev12,stability} we have introduced the concept of embedding which enables to replace the infinite system (central region + left and right electrodes) by a finite open system whose
reduced density matrix is governed by the Lindblad master equation.
The main idea is to divide each electrode  into two parts: an infinite environment and a sufficiently large but finite buffer between the central region and the environment.
Then, projecting out the  environment degrees of freedom and employing the Born-Markov approximation,  we obtain the Lindblad master equation~\eqref{lindblad} for
the density matrix of the embedded system (central region + left and right buffers, see Fig.~\ref{system}).  The
 Hamiltonian of the embedded system has the form
\begin{equation}
  H = H_S + H_{T}+H_B,
\end{equation}
where $H_{B}=\sum\limits_{k\alpha}\varepsilon_{k\alpha}a^\dag_{k\alpha} a_{k\alpha}$ describes the left ($\alpha=L$) and right ($\alpha =R$) finite
buffers ($k=1,\ldots,N$), and $H_{T}=\sum\limits_{k\alpha}t_{k\alpha}(a^\dag_{k\alpha}\alpha + \mathrm{h.c.})$ is the tunneling  coupling between the central region and the buffers.
The non-Hermitian  part of the master equation is
  \begin{align}\label{non_herm}
{\Pi}\rho(t) = \sum_{k\alpha} \sum_{\mu=1,2} \bigl(2 L_{ k\alpha   \mu}\rho(t) L^\dag_{ k\alpha  \mu} - \{L^\dag_{k \alpha \mu} L_{ k \alpha  \mu},\rho(t) \}\bigr),
\end{align}
where
\begin{align}\label{L_operators}
  L_{k\alpha 1} = \sqrt{\Gamma_{k\alpha 1}}a_{k \alpha},~~L_{k \alpha  2} = \sqrt{\Gamma_{k\alpha 2}  }a^\dag_{ k  \alpha}.
\end{align}
Here $\Gamma_{k\alpha 1}=\gamma_{k\alpha}(1-f_{k\alpha})$,  $\Gamma_{k\alpha 2}=\gamma_{k\alpha}f_{k\alpha}$,  $f_{k\alpha} =[1+ e^{(\varepsilon_{k\alpha} - \mu_\alpha)/T }]^{-1}$
and $\gamma_{k\alpha} $ is determined through the imaginary   part of the self-energy arising from the buffer - environment interaction.

Using the formalism of superoperators described in section 2, we rewrite the obtained master equation in the Schr\"{o}dinger-like form ~\eqref{Schrodinger}.
The Liouvillian of the embedded system is given by
\begin{equation}\label{L_el-ph}
L= L_0 + L_T + L_\kappa.
\end{equation}
Here, $L_0=L_\mathrm{el}+L_\mathrm{ph}+L_B$ describes the uncorrelated system,
$L_T=\hat H_T-\widetilde H_T$, and
\begin{equation}
  L_\kappa = \kappa\{\hat\alpha^\dag\hat\alpha(\hat d^\dag + \hat d)-(\mathrm{t.c.})\bigr\}
\end{equation}
represents the electron-phonon correlations. Hereinafter,
the notation '(t.c.)' stands for items which are tilde-conjugated (see the tilde-cojugation rules in section~2) to displayed ones. The steady-state current through the central region
is given by
\begin{equation}\label{currentSO}
  J_\alpha = i\sum_k t_{k\alpha} \braket{I|\hat a^\dag_{k\alpha}\hat\alpha - \hat \alpha^\dag\hat a_{k\alpha}|\rho},
\end{equation}
which requires the solution of Eq.~\eqref{StSt_def} with  Liouvillian~\eqref{L_el-ph}.

To solve Eq.~\eqref{StSt_def} within the NECC method,  we should first define an appropriate reference state.
For the problem under consideration, as the reference state we take the density matrix $\ket{\rho_0}$ of the  uncorrelated system, i.e.,
\begin{equation}\label{rho 0}
L_0\ket{\rho_0}=0.
\end{equation}
In  Appendix~A we demonstrate how to introduce normal modes of $L_0$, such that $\bra{I}$ and $\ket{\rho_0}$ would be the vacuum states for
respective creation and annihilation superoperators.  In terms of these uncorrelated superoperators $L_0$ is diagonal
\begin{equation}
 L_0 = \varepsilon_0\hat\beta^\dag\hat\beta+\omega_0\hat\gamma^\dag\hat\gamma+ \sum_{k\alpha}E_{k\alpha}\hat b^\dag_{k\alpha}\hat b_{k\alpha} - (\mathrm{t.c.}),
\end{equation}
 while  the parts of the
Liouvillian responsible for electron-phonon correlations and the coupling between the central region and the buffers read as
 \begin{align}\label{el-ph}
  L_\kappa=\kappa\Bigl\{\hat\beta^\dag\hat\beta\bigl[(1+N_{\omega})\hat\gamma^\dag + N_\omega\widetilde\gamma^\dag + \hat\gamma+\widetilde\gamma\bigr]
  + i\widetilde\beta\hat\beta\hat\gamma^\dag - (\mathrm{t.c.})\Bigr\}
\end{align}
and
 \begin{equation}\label{L_B-el}
   L_T=-\sum_{k\alpha} t_{k\alpha}\bigl\{\hat\beta^\dag \hat b_{k\alpha}+\hat b^\dag_{k\alpha}\hat\beta  - if_{k\alpha}\hat\beta^\dag\widetilde b^\dag_{k\alpha}-(\mathrm{t.c.})\bigr\},
 \end{equation}
respectively. In~\eqref{el-ph}, $N_\omega$ denotes the number of equilibrium thermally excited phonons (see Appendix~A).

Due to interaction terms~(\ref{el-ph}, \ref{L_B-el}) the correlated steady-state density matrix contains multiconfigurational excitations above the reference state  $\ket{\rho_0}$.
For the considered problem the multiconfigurational creation superoperators $C^\dag_i$ are product of bosonic and fermionic creation superoperators. More specifically
the index $i$ becomes the set of occupation numbers,
\begin{align}
  i\to~ &(n^i_{\hat\gamma},n^i_{\widetilde\gamma}\,;n^i_{\hat\beta},n^i_{\widetilde\beta}\,;n^i_{\hat b_{L1}},\ldots,n^i_{\hat b_{LN}},n^i_{\widetilde b_{L1}},\ldots,n^i_{\widetilde b_{LN}}
 ;n^i_{\hat b_{R1}},\ldots,n^I_{\hat b_{RN}},n^i_{\widetilde b_{R1}},\ldots,n^i_{\widetilde b_{RN}}),
 \notag\\
 &n^i_{\hat\gamma},n^i_{\widetilde\gamma}=0,1,2,\ldots,~~~
 n^i_{\hat\beta},\,n^i_{\widetilde\beta}=0,1,~~\mathrm{and}~~
 n^i_{\hat b_{\alpha k}},\, n^i_{\widetilde b_{\alpha k}} =0,1,~~~\forall \alpha,\,k,
\end{align}
and the generic multiconfigurational creation superoperators $C^\dag_i$ is given by
\begin{equation}
  C^\dag_i\to (\hat\gamma^\dag)^{n^i_{\hat\gamma}} (\widetilde\gamma^\dag)^{n^i_{\widetilde\gamma}}
  (\hat\beta^\dag)^{n^i_{\hat\beta}}(\widetilde\beta^\dag)^{n^i_{\widetilde\beta}}
  \prod_{\alpha k} (\hat b^\dag_{\alpha k})^{n^i_{\hat b_{\alpha k}}}\prod_{\alpha k} (\widetilde b^\dag_{\alpha k})^{n^i_{\widetilde b_{\alpha k}}}.
\end{equation}
Moreover, from the structure of $L_\kappa$ and $L_T$ it follows that in each $C^\dag_i$ the number of non-tilde fermionic superoperators should be equal to the number of tilde ones, i.e.,
\begin{equation}
 n^i_{\hat\beta} + \sum_{\alpha k} n^i_{\hat b_{\alpha k}} = n^i_{\widetilde\beta} + \sum_{\alpha k} n^i_{\widetilde b_{\alpha k}},~~~\forall i.
\end{equation}

Due to the phonon subsystem the number of possible multiconfigurational excitations is infinite.
Furthermore, since $L_\kappa$  involves terms with two annihilation superoperators the  expansion~\eqref{series} terminates after the second order commutators.
Therefore, for the considered problem, the exact cluster superoperator $S$ is given by the solution of the infinite system of  coupled quadratic nonlinear equations.

In the present work we restrict our consideration by the cluster correlation superoperator $S_1$ and $S_2$ which includes
only terms linear with respect to
phonon creation superoperators. The superoperator $S_1$ is quadratic in terms of fermion creation superoperators
\begin{align}\label{Sprime}
  S_1=&W(\hat\gamma^\dag+\widetilde\gamma^\dag)-i \hat\beta^\dag\widetilde \beta^\dag (n+n_{10}\hat\gamma^\dag+n_{01}\widetilde\gamma^\dag)
  \notag\\
  &+ i\sum_k\bigl\{\hat\beta^\dag\widetilde b^\dag_{k}(I_k+I_{k10}\hat\gamma^\dag+I_{k01}\widetilde\gamma^\dag) - (\mathrm{t.c.})\bigr\}
  \notag\\
  &+i\sum_{kl}\hat b^\dag_k\widetilde b^\dag_l(F_{kl}+F_{kl10}\hat\gamma^\dag+F_{kl01}\widetilde\gamma^\dag)~,
  \end{align}
  while $S_2= S_1+S'$ contains four-fermion components
\begin{equation}\label{Sp}
  S'= + i\sum_{kl}\hat\beta^\dag\widetilde\beta^\dag\hat b^\dag_k\widetilde\beta^\dag_l(G_{kl}+G_{kl01}\hat\gamma^\dag + G_{kl10}\widetilde\gamma^\dag)
\end{equation}
Here, to simplify the notations, we omit the index $\alpha=L,R$ implying that the summation is performed over the states in the left and right buffers, i.e., $k,l\in L,R$.
Substituting $S_1$ or $S_2$ into~\eqref{series}  we obtain
the  set of nonlinear (quadratic) equations for the cluster amplitudes $n$, $W$, $I$, $F$, and $G$~(see Eq.~\eqref{system2} in Appendix~B). The solution of these equations
 provides an approximate steady-state density matrix for the considered nonequilibrium charge transport problem. From $S_{1,2}=(S_{1,2})\widetilde{}$~ it follows that $n$ and $W$ are real, while
$n^*_{10}=n_{01}$, $F^*_{kl}=F_{lk}$, $F^*_{kl10}=F_{lk01}$, $G^*_{kl}=G_{lk}$, and $G^*_{kl10}=G_{lk01}$. In what follows, to distinguish between results obtained with $S_1$ and $S_2$
superoperators, we will refer to them as NECC(1) and NECC(2), respectively.   

To compute the steady-state current we express the current superoperator in Eq.~\eqref{currentSO} in terms of uncorrelated creation and annihilation superoperators. We get
\begin{align}
 J_\alpha&=\sum_{k\in\alpha} t_{k} \braket{I|\widetilde b_{k}\hat\beta - \widetilde\beta \hat b_{k}|\rho}
=-2\mathrm{Im}\sum_{k\in\alpha} t_{k}I_{k}.
\end{align}
Thus, the steady-state current is expressed through the cluster amplitudes $I_{k}$. On the other hand, with the above relation the first equation in~\eqref{system2} becomes
\begin{equation}
  J_L+J_R=0.
\end{equation}
Therefore, we can say that the first equation in~\eqref{system2} guaranties the current conservation.
Moreover, it can be easily shown
that this equation is exact in the sense that the inclusion
of extra terms in $S_{1,2}$ does not modify it.  Consequently, the presented coupled cluster approach for the charge transport problem is a current conserving one.

\subsection{Numerical calculations}

 In our numerical calculations the left and right electrodes are represented by two semi-infinite  tight-binding chains of sites.
The coupling between the central region and the edge sites is given by the matrix elements~$t_\alpha$ ($\alpha=L,~R$).
Each electrode is characterized by the hopping matrix element~$h_\alpha$ and the on-site energy~$\epsilon_\alpha$. We choose the
following  parameters: $h_L=h_R=2.5$, $t_L=t_R=1.0$ and the temperature $T=0.1$.
Additionally, we assume that the electrodes are half filled, i.e., the corresponding left and right chemical potentials are positioned at $\epsilon_{L,R}$, and
the applied voltage $V=1.0$ symmetrically shifts the on-site energies, $\epsilon_{L,R} = \pm 0.5V$.

For the embedded system we take $N=800$ atoms from each electrode as buffers.
This size of the buffers has been proven to give the exact results for the  steady-state current calculated within the mean-field approximation
and the second order perturbation (SOPT) theory ~\cite{dzhioev11a,dzhioev11b,dzhioev12}. In~\cite{dzhioev14} we  justify this choice of $N$ for nonequilibrium configuration
interaction (NECI) calculations. Following~\cite{dzhioev14}, we  studied the convergency of NECC results against increasing the size of buffers.
We found that when increasing $N$ the NECC current converges to some limit value. Moreover, for $N\ge 800$ the results are affected only marginally by
increase of $N$ regardless of the particular choices of model parameters.


\begin{figure*}[t]
 \begin{centering}
\includegraphics[width=1.0\textwidth]{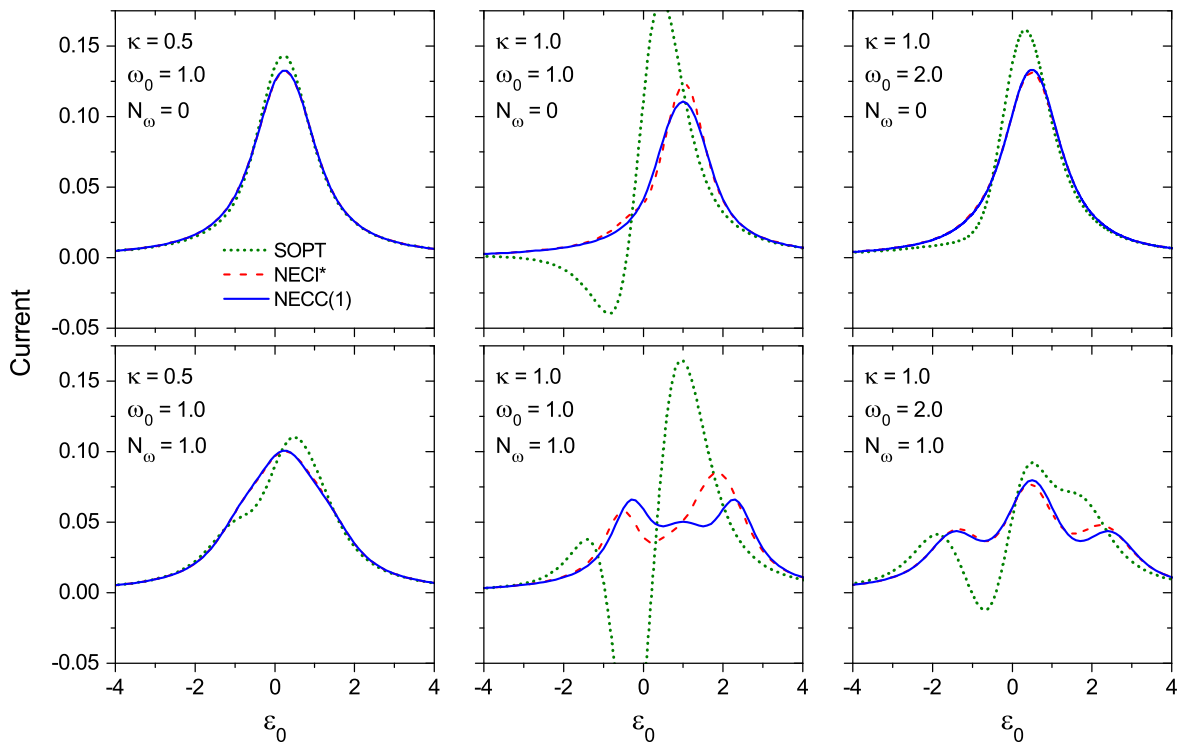}
\caption{Current through the central region calculated within different approaches as a function  of the level energy, $\varepsilon_0$. The calculated NECC(1) currents are shown
  along with the currents obtained within the second order perturbation theory (SOPT)~\cite{dzhioev12} and the nonequilibrium configuration interaction method (NECI$^*$)~\cite{dzhioev14}. }
 \label{fig2}
 \end{centering}
\end{figure*}

Before considering obtained results, let us briefly discuss numerical solution of the system of equations~\eqref{system2} for cluster amplitudes.
The set of nonlinear equations is solved by applying Newton's iterative scheme. 
It is clear that the main computational cost comes from large buffer zones. Namely, for the cluster correlation superoperator quadratic with respect to $\hat b^\dag$,~ 
$\widetilde b^\dag$ the dimension of the system increases as $4N^2$. However the system is sparse and number of nonzero elements is proportional to $4N^2$. For
for $N=800$ the system can be solved on a standard personal computer (e.g., 16~GB of RAM and Intel Core i5 quad-core CPU with 3.4~GHz clock speed) in about one hour.
To solve the system we have used  sparse solver routines from Intel MKL Fortran library.

To illustrate the approach, we compute the current  through the central region as a function of the level energy $\varepsilon_0$. In figure~\ref{fig2}, each panel displays
the current $J_L$ calculated for a particular choice of model parameters.
We compare, for the same sets of model parameters,  the calculated NECC(1) currents with those obtained by the second order perturbation theory~\cite{dzhioev12} and with
the nonequilibrium configuration interaction method built on the coherent reference state (a so-called NECI$^*$ approximation) ~\cite{dzhioev14}.

In the upper panels of figure~\ref{fig2} we consider the case when $N_\omega=0$, that means that the vibrational state in the system is not thermally excited.
In this case the NECC(1) current reaches a maximum value exactly at $\varepsilon_0=\kappa^2/\omega_0$. It is the so-called reorganization
energy~\footnote{This reorganization energy is also referred to  as polaron-shift~\cite{lang_firsov1963,mahan2000}.} of an
electronic state associated to the electron-phonon coupling.
Therefore, when $\varepsilon'_0=\varepsilon_0 - \kappa^2/\omega_0$ is close to chemical potentials of the electrodes (resonant regime) the NECC(1) current reaches its
maximum value.

From the upper-left panel we see that for a weak electron-phonon coupling ($\kappa=0.5$) all three approaches predict the currents which closely follow each
other for all values of $\varepsilon_0$.
This resemblance vanishes, however, with an increase in the coupling strength (see the upper-middle panel):  In the resonant regimes the SOPT and  NECI$^*$ currents
overestimate the NECC result and, in addition,   the SOPT current becomes unphysical negative in the off-resonant regime.
The negative SOPT current disappears if we consider the case of larger phonon energy (see the upper-right panel).
For this case, the NECC and NECI$^*$ calculations give the same result, while the maximum value of the SOPT current noticeably  exceeds the maximum values of the NECC(1) and NECI$^*$  currents.

The advantages of the NECC method become more apparent by considering the case when the central region contains a nonzero number of
thermally excited equilibrium vibrational quanta. For $N_\omega=1.0$, the results of such calculations are shown in the lower panels of figure~\ref{fig2}. As obvious from the
graphs the NECC(1) current is symmetric with respect to $\varepsilon_0 = \kappa^2/\omega_0$. For the identical electrodes employed
 this result is a consequence of a particle-hole symmetry of the whole system.  Indeed, by applying the particle-hole transformation to all fermionic operators
and then performing a shift of the phonon field we get the same system but with an electronic state with energy $\varepsilon'_0 =  - \varepsilon_0 + 2\kappa^2/\omega_0$.
Therefore, the current should be symmetric with respect to  $\varepsilon_0 = \kappa^2/\omega_0$.

Contrary to the NECC method,  the SOPT and NECI$^*$ approaches do not fulfill the symmetry property.
This is  most clearly evident in the lower-middle panel of  figure~\ref{fig2}, where both the SOPT and NECI$^*$ approaches demonstrate the substantially asymmetric behaviour.
It is interesting to note, the behaviour of the NECI$^*$ current becomes more symmetric when we decrease the coupling strength or increase the phonon energy. As a result,
for $\kappa=0.5$ the difference between the NECC(1) and  NECI$^*$ currents is marginally negligible, while for $\kappa=1.0$
and $\omega_0=2.0$ both the approaches give results close to each other.

\begin{figure*}[t]
 \begin{centering}
\includegraphics[width=1.0\textwidth]{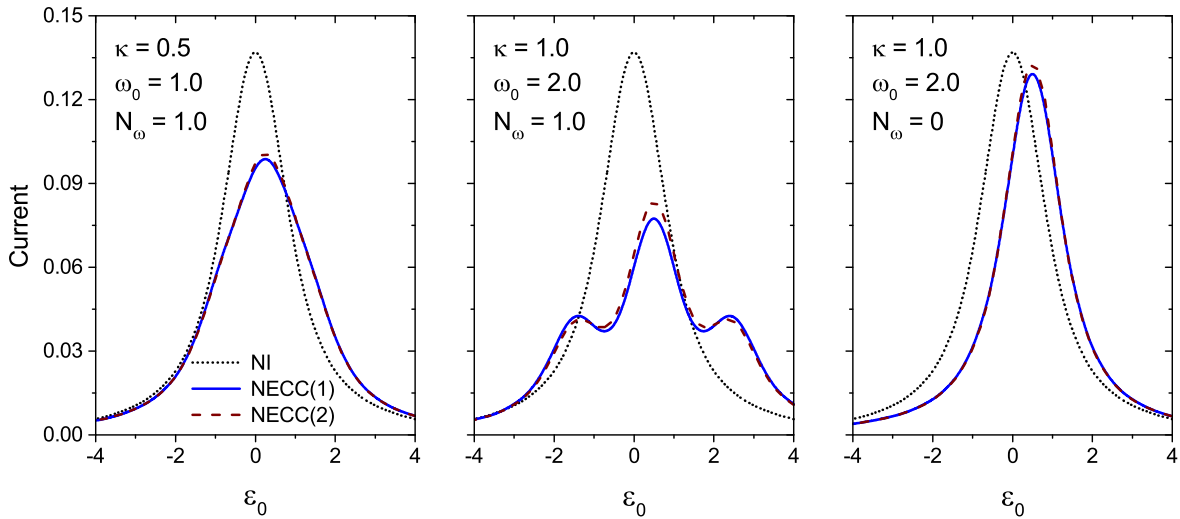}
\caption{Current through the central region calculated within NECC(1) and NECC(2)  approaches as a function  of the level energy, $\varepsilon_0$.
 For comparison purpose, the current through noninteracting (NI) level is also shown.}
 \label{fig3}
 \end{centering}
\end{figure*}

In Fig.~\ref{fig3} we compare NECC(1) and NECC(2) electron currents. We also show the current  for the noninteracting ($\kappa = 0$) electronic level for the comparison in this figure. One can see from  the figure,
the inclusion of higher order term $S'$ (see Eq.~\eqref{Sp}) into cluster correlation superoperator does not change substantially the results of the NECC(1) approach.  
The difference between the NECC(1) and  NECC(2) currents is negligible for weak coupling (left panel).  Moreover, 
the inclusion of four-fermion
component into cluster correlation superoperator leaves the $\varepsilon_0$ dependence of the current almost unchanged even in the strong coupling  regime (middle and right panels) -- the difference between NECC(1) and NECC(2)  is only a few percent of the total current.  Thus, we can say that even the use of excited cluster superoperator $S_1$, which is quadratic in in fermion creation superoperators (\ref{Sprime}), captures the most of nonequilibrium correlations in our model.

\section{Conclusions}

In this paper, we propose an extension of the coupled cluster method to nonequilibrium correlated many-body quantum systems described by the
Lindblad master equations. The method is based on a superoperator representation of the master equation.
It was shown that in Liouville-Fock space  the steady-state density matrix can be defined via the exponential ansatz $\ket{\rho}=\exp(S)\ket{\rho_0}$, where
$\ket{\rho_0}$ is a reference state and $S$ is a cluster correlation superoperator linearly expanded in terms of a complete set of multiconfigurational creation superoperators.
A general prescription how to calculate expansion (cluster) amplitudes was given.

As an application of the method, quantum transport through a vibrational electronic level was considered. To compute the steady-state current a truncated
expansion of the cluster correlation superoperator was used. It was shown that the method always preserves the continuity equation, no matter how we truncate the cluster correlation superoperator.
It was demonstrated that the NECC method provides advantages over the perturbation theory and the nonequilibrium configuration interaction method when the electron-phonon
coupling is large or when the system contains thermally excited vibrational quanta.

To conclude we would like to mention some possible further steps. An obvious application of the NECC method is to
apply it to the charge transport problem through the system with electron-electron correlations.   Another route is to extend the method to a dynamical nonequilibrium case by making
the cluster amplitudes time-dependent functions.  All these will be the subject of our future investigations.

\appendix

\section{Uncorrelated density matrix  as a vacuum state}

Here we demonstrate how to define normal modes for the Liouvillian $L_0=L_\mathrm{el}+L_\mathrm{ph}+L_B$, where
\begin{align}
  L_\mathrm{el}=&\varepsilon_0(\hat\alpha^\dag\hat\alpha - \widetilde \alpha^\dag\widetilde\alpha),~~~
  L_\mathrm{ph}= \omega_0(\hat d^\dag\hat d - \widetilde d^\dag\widetilde d),
  \notag\\
  L_\mathrm{B}=&\sum_{k\alpha}\bigl\{\varepsilon_{k\alpha}(\hat a^\dag_{k\alpha}\hat a_{k\alpha}-\widetilde a^\dag_{k\alpha}\widetilde a_{k\alpha})-i\Pi_{k\alpha}\bigr\},
  \notag\\
  \Pi_{k\alpha}=&(\Gamma_{k\alpha1}-\Gamma_{k\alpha2})(\hat a^\dag_{k\alpha}\hat a_{k\alpha}+
  \widetilde a^\dag_{k\alpha}\widetilde a_{k\alpha}) \notag\\&-
  2i(\Gamma_{k\alpha1}\widetilde a_{k\alpha}\hat a_{k \alpha}+\Gamma_{k\alpha2}\widetilde a^\dag_{k\alpha}\hat a^\dag_{k\alpha})+2\Gamma_{k\alpha2}
\end{align}
 such that the uncorrelated  density matrix
$\ket{\rho_0}=\ket{\rho}_\mathrm{el}\ket{\rho}_\mathrm{ph}\ket{\rho}_\mathrm{B}$ and the super-vector $\bra{I}$ would be the right and left  vacuum states, respectivelly.

First of all we note that due to  Eq.~\eqref{ketA} the superoperators
\begin{align}\label{creation1}
  \hat\beta^\dag = \hat \alpha^\dag - i\widetilde\alpha,~~
  \hat\gamma^\dag = \hat d^\dag -\widetilde d,~~~
  \hat b_{k\alpha}^\dag = \hat a^\dag_{k\alpha} - i\widetilde a_{k\alpha}
\end{align}
and their tilde conjugated partners $\widetilde\beta^\dag$, $\widetilde\gamma^\dag$, $\widetilde b_{k\alpha}^\dag$ annihilate $\bra{I}$ when acting from the right.
However, since $\bra{I}$ and $\ket{\rho_0}$ are not mutually Hermitian conjugate, the respective annihilation superoperators can not be defined by taking the Hermitian
 conjugate~of~(\ref{creation1}).

To define annihilation superoperators associated with the central region we assume that that does not contain electrons when  disconnected from the electrodes,
while there are a certain number $N_\omega$ of thermally excited equilibrium vibrational quanta
\begin{equation}
  \braket{I|\hat\alpha^\dag\hat\alpha|\rho}_\mathrm{el}=0,~~~~ \braket{I|\hat d^\dag\hat d|\rho}_\mathrm{ph}=N_\omega.
\end{equation}
Then we introduce
\begin{align}\label{beta_gamma}
    \hat\beta =  \hat a,~~\widetilde\beta=(\hat\beta)\widetilde{},~~\hat \gamma =(1+N_\omega)\hat d - N_\omega \widetilde d^\dag,~~\widetilde\gamma=(\hat\gamma)\widetilde{}
 \end{align}
such that $\hat \beta\ket{\rho}_\mathrm{el} = \widetilde \beta\ket{\rho}_\mathrm{el}=0$, $\hat\gamma\ket{\rho}_\mathrm{ph} = \widetilde\gamma\ket{\rho}_\mathrm{ph}=0$
and
\begin{align}
 L_\mathrm{el} = \varepsilon_0(\hat\beta^\dag\hat\beta - \widetilde\beta^\dag\widetilde\beta ),~~
  L_\mathrm{ph} = &\omega_0(\hat\gamma^\dag\hat\gamma - \widetilde\gamma^\dag\widetilde\gamma ).
\end{align}
As for the buffers, we can apply the equation of motion method and  define
\begin{equation}\label{b}
  \hat b_{k\alpha} =  (1-f_{k\alpha})\hat a_{k\alpha} - i f_{k\alpha} \widetilde a^\dag_{k\alpha},~~
  \widetilde b_{k\alpha} = (\hat b_{k\alpha})\widetilde{}
\end{equation}
such that
\begin{equation}
 L_\mathrm{B} = \sum_{k\alpha} (E_{k\alpha} \hat b^\dag_{k\alpha} \hat b_{k\alpha} - E^*_{k\alpha} \widetilde b^\dag_{k\alpha} \widetilde b_{k\alpha}),
 \end{equation}
where $E_{k\alpha}=\varepsilon_{k\alpha}-i\gamma_{k\alpha}$. Thus, we have $\hat b_{k\alpha}\ket{\rho}_\mathrm{B} = \widetilde b_{k\alpha}\ket{\rho}_\mathrm{B}=0$.

It should be recognized that although the above defined uncorrelated creation and annihilation operators are not Hermitian conjugate to each other,
they satisfy the canonical (anti)commutation relations.

\section{System of nonlinear equations for coupled cluster amplitudes}

The system of nonlinear equations for the amplitudes in the cluster correlation operator~$S_2$ has the following form
\begin{align}\label{system2}
  &\sum_k t_k(I_k-I^*_k)=0;
  \notag\\
  &W\omega_0 + \kappa n=0;
  \notag\\
  &n_{10}\omega_0 - \sum_k t_k(I_{k10}-I^*_{k01}) + \kappa n(1-n)=0;
  \notag\\
  &I_k\bigl(\varepsilon_0 - E^*_k + 2\kappa W) -t_k n - \sum_l t_l F_{lk}+\kappa(I_{k10}+I_{k01}) = -t_k f_k;
  \notag\\
  &I_{k10}\bigl(\varepsilon_0 + \omega_0 - E^*_k + 2\kappa W\bigr) -t_k n_{10} + \kappa I_k(1-n+N_\omega)-\sum_l t_l F_{lk10}=0;
   \notag\\
   &I_{k01}\bigl(\varepsilon_0 - \omega_0 - E^*_k + 2\kappa W\bigr) -t_k n_{01} + \kappa I_k(n+N_\omega)-\sum_l t_l F_{lk01}=0;
   \notag\\
  &F_{kl}(E_k-E^*_l)-t_kI_l+t_lI^*_k=0;
   \notag\\
  &F_{kl10}(E_k-E^*_l+\omega_0)-t_kI_l+t_lI^*_k+\kappa I^*_kI_l + \kappa G_{kl}=0;
  \notag\\
  &G_{kl}(E_k-E^*_l) + \kappa I_l(I^*_{k10}+I^*_{k01})-\kappa I^*_k(I_{l10}+I_{l01})=0;
  \notag\\
  &G_{kl10}(E_k-E^*_l+\omega_0)+\kappa G_{kl}(1-2n)+\kappa(I^*_{k10} I_{l10} -I^*_{k01} I_{l01})=0.
    \end{align}
To obtain equations for the cluster amplitudes in $S_1$ we should neglect the last two equations and put $G_{kl}=0$.

In the above equations, the nonequilibrium boundary conditions are introduced via fermionic occupation numbers $f_k$
which depend on the chemical potentials in the left ($k\in L$) and right electrodes ($k\in R$).
Neglecting the electron-phonon correlations $(\kappa=0)$ we get the system of linear equations
for the cluster amplitudes $n,~I_k,~F_{kl}$ which determine the exact steady-state density matrix for the charge transport problem through
a noninteracting level.

\section*{Reference}


\end{document}